\documentclass[aps,prl,twocolumn]{revtex4}

\usepackage{amsmath,amssymb,graphicx,color}

\newcommand{\eqn}{\begin{equation}}
\newcommand{\nqe}{\end{equation}}

\newcommand{\bra}[1]{\langle #1 \rvert}
\newcommand{\ket}[1]{\lvert #1 \rangle}

\DeclareMathOperator{\sgn}{sgn}

\DeclareMathOperator{\erf}{erf}

\DeclareMathOperator{\erfc}{erfc}

\begin{document}

\title{Quench dynamics of the interacting Bose gas in one dimension}
\author{Deepak  Iyer} \author{Natan Andrei} \affiliation{Department of Physics, Rutgers University\\
  Piscataway, New Jersey 08854.}  \date{\today}

\begin{abstract}
  We obtain an exact expression for the time evolution of the
  interacting Bose gas following a
  quench from a generic initial state using the Yudson representation
  for integrable systems.  We study the time evolution  of the
  density and noise correlation for a small number of bosons and
  their asymptotic behavior for any number. We show that for
  any value of the coupling, as long as it is repulsive, the system asymptotes towards a
  strongly repulsive gas, while for any value of an attractive coupling
  the long time behavior is dominated by the maximal bound state. This occurs
  independently of the initial state and
  can be viewed as an emerging ``dynamic universality''.
\end{abstract}

\maketitle

The interest in quantum systems out of equilibrium has been re-ignited
with the advent of ultra-cooled atoms in laser traps
or lithographic nano-devices, which allow finely controlled experiments
to explore various aspects of non-equilibrium dynamics -- quenching or
non-linear transport being examples
\cite{blochdali,gorlitz,moritz,kinoshita}. In parallel, advances in theoretical
computational techniques now allow us to calculate the time evolution of various
observables of interacting many body systems and study the underlying
physics~\cite{rigol1,rigol2}. Many of the systems studied are
governed, to a very close approximation, by integrable Hamiltonians
~\cite{kinoshita, cazalilla}, opening the way to understanding their quantum
dynamics in great detail.

We shall be concerned here with non-equilibrium \emph{quench} dynamics
-- preparing the system in some initial state $|\Phi_0\rangle$ and
following its evolution under the influence of a Hamiltonian $H$
applied to it suddenly at time $t=0$. For example, a Bose
gas initially in a Mott state in a deep periodic trap, is time evolved
under a superfluid Hamiltonian with the trap depth reduced,
allowing the bosons to easily hop from site to site. The
evolution can be formally obtained by expanding the initial state in
terms of a complete set of orthonormal energy eigenstates
$\{\lambda\}$,
\begin{equation}
  \label{eq:timeevol}
  \begin{split}
    \ket{\Phi_0,t} &= e^{-iHt}\ket{\Phi_0}
    =\sum_{\{\lambda\}}e^{-iE_\lambda
      t}\ket{\lambda}\bra{\lambda}\Phi_0 \rangle.
  \end{split}
\end{equation}
For an exactly solvable model, these eigenstates $\ket{\lambda}$ are
given by the Bethe Ansatz. This approach has been
effective in obtaining the full spectrum (and thermodynamics) of a
number of important models~\cite{lieblin, andrei, tsvelik, takahashibk}.  However, for
time evolution, in addition to the spectrum, one also needs to
compute the overlap of the eigenstates with the initial state and
carry out the summation -- difficult tasks due to the complexity of
the Bethe Ansatz eigenstates. In spite of this, some progress has been
made~\cite{caux,calabrese,gritsev}.

In this letter, we address this question generalizing a ``contour
integral'' approach introduced by V.~Yudson for the purpose of
studying superradiance effects in the context of the Dicke model
~\cite{yudson1,yudson2}. The contour representation does away with
both the above steps, and replaces the summation over eigenvalues with
an integral along contours in the complex plane, appropriately chosen
to faithfully represent the initial state, capturing the relevant
overlaps via residues of poles in the two particle S-matrix of the Bethe Ansatz
eigenstate.

We show that the framework has general applicability and use
it to understand the quench dynamics of an interacting gas of bosons,
with short range attractive or repulsive interactions,
\begin{equation}
  \label{eq:llham}
  H = \int_x \partial b^\dag(x) \partial b(x) + c b^\dag(x)b(x)b^\dag(x)b(x).
\end{equation}
We shall obtain the time evolution of an arbitrary initial state, and
calculate the evolution of the density and noise correlations.

The model~\eqref{eq:llham} was solved by Lieb and Liniger~\cite{lieblin}. Its
$N$-particle eigenstates, labeled by the momenta $\{\lambda\}$, are:
\begin{equation}
  \label{eq:lleigen}
  \ket{\lambda} = \int_y \prod_{i<j}\mathcal{S}_y\big(Z^y_{ij}(\lambda_i-\lambda_j)
  \prod_j e^{i\lambda_j y_j}\big)b^\dag(y_j)\ket{0},
\end{equation}
where $\mathcal{S}_y$ is a symmetrizer. The factor, $Z^y_{ij}(z) =
\frac{z-ic\sgn(y_i-y_j)}{z-ic}$, incorporates the $S$-matrix, $S_{ij}(\lambda_i-\lambda_j)=
\frac{\lambda_i- \lambda_j+ic}{\lambda_i - \lambda_j-ic}$, that
describes the scattering of two bosons with momenta $\lambda_i,
\lambda_j$. The corresponding eigenenergies are $E_\lambda=
\sum_{j=1}^N \lambda_j^2$ with the momenta being real-valued for repulsion
($c>0$), or complex conjugate pairs (signifying bound states) for attraction.
One usually proceeds by imposing periodic boundary conditions to
determine the values of the $\{\lambda \}$'s and subsequently studies
the infinite volume limit.  Instead, we shall work directly in the
infinite volume limit with the momenta $\{\lambda \}$ unconstrained,
allowing us to integrate over them instead of summing over discrete
values.

A given state $ |\Phi_0\rangle= \int_x  \,\Phi_0(x_1 \cdots
x_N) \prod_{j=1}^N b^{\dagger}(x_j) |0\rangle$ ($\Phi_0$ symmetric) can be directly time evolved if it
is expressed in terms of the eigenstates, eq.~\eqref{eq:lleigen}.
It can be shown that, (we denote $\theta(x_1>\cdots>x_N) =\theta(\vec{x})$),
\begin{equation}
  \begin{split}
    \label{eq:claim}
    |\Phi_0\rangle &=  \int_{x,y,\lambda}\, \theta(\vec{x}) \Phi_0(\vec{x})
    \big(\mathcal{S}_y \prod_j e^{i\lambda_j(y_j-x_j)} \\
    &\qquad\times \prod_{i<j}
    Z^y_{ij}(\lambda_i-\lambda_j)\big) b^\dag(y_j)\ket{0}
  \end{split}
\end{equation}
where the momenta integration contours are determined by
the interaction. In the repulsive case, all
$\{\lambda_j \}$ have to be integrated along the real line, while in the
attractive case, the $\lambda_j$ have to be integrated along lines
parallel to the real axis that are separated in the imaginary direction
by a little more than $|c|$. This separation captures the bound
states that appear in the spectrum of this model.  The proof of the statement involves
studying ordered initial states, $|\vec{x}\rangle
=\theta(\vec{x}) \prod_j b^\dag(x_j)\ket{0}$ which, when represented
in the form above, have zero residues for the chosen integration
contours. It can then be shown that each $\lambda_j$ integration
reduces to $\delta(y_j-x_j)$ proving the validity of \eqref{eq:claim}.
For details of the formalism as applied to the interacting Bose gas, please see~\cite{deebig}.


In the rest of this article, we discuss the quenching of a system of
bosons initially in a Mott state (a deep periodic trap) or in a condensate
(a parabolic trap) with the trap removed at $t=0$ and the system evolving
subsequently under the action of the interacting Hamiltonian~\eqref{eq:llham}.
We shall be interested in the effects of
interaction, attractive and repulsive, on the evolution.

\begin{figure}[htb]
  \centering \fbox{\includegraphics[width=6cm]{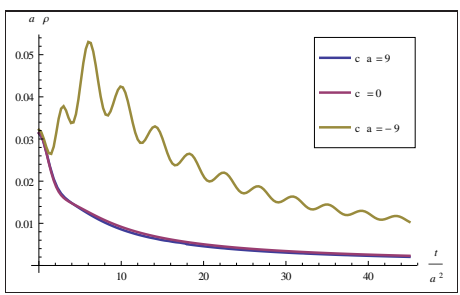}}
  \caption{$\langle\rho(x=0,t)\rangle$ vs. $t$, after the
    quench from $\ket{\phi_{\text{latt}}}$. $\sigma/a\sim 0.3$}
  \label{fig:twores}
\end{figure}

\begin{figure}[htb]
  \centering \fbox{\includegraphics[width=6cm]{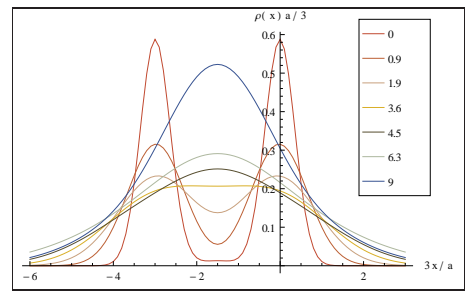}}
  \caption{$\langle \rho(x,t)\rangle$ vs. $x$ of two attractive bosons
    after a quench from $\ket{\phi_{\text{latt}}}$ plotted for different times. The two separate peaks coalesce
    and form a bound state as $t\to\infty$. The key shows the times (in units of $a^2$).}
  \label{fig:wfatt2}
\end{figure}

We first discuss the case of bosons trapped in a periodic potential $ \ket{\Phi_{\text{latt}}} =
\prod_j\frac{1}{(2\pi\sigma^2)^{\frac14}}
\int_xe^{-\frac{(x_j+(j-1)a)^2}{2\sigma^2}}b^\dag(x_j)\ket{0}$.  If we assume that
the wave functions of neighboring bosons do not overlap significantly,
i.e., $e^{-\frac{a^2}{\sigma^2}}\ll 1$, then an ordering of the
initial particles (required in the Yudson representation), with
$x_j=(1-j)a$, is induced by the non-overlapping support. To illustrate this,
we find the exact two particle finite time wave function:
\begin{equation*}
  \label{eq:rep2p}
  \begin{split}
    & \ket{\Phi_{\text{latt}},t}_2  = \frac{1}{4\pi i t}\int_{x,y}e^{-\frac{x_1^2}{\sigma^2}-\frac{(x_2+a)^2}{\sigma^2}} e^{i\frac{(y_1-x_1)^2}{4t}+i\frac{(y_2-x_2)^2}{4t}}\\
    & \left[1- c\sqrt{\pi i
        t}\theta(y_2-y_1)e^{\frac{i}{8t}\alpha(t)^2}
      \erf\left(\frac{i-1}{4}\frac{i\alpha(t)}{\sqrt{t}}
      \right)\right] b_{y_1}^\dag b_{y_2}^\dag\ket{0},
  \end{split}
\end{equation*}
where $\alpha(t) = 2ct-i(y_1-x_1)-i(y_2-x_2)$. In the attractive case
with the same initial conditions we find the same result with $\erfc$
(the complementary error function) replacing $\erf$. This induces a
significant change in behavior, as can be seen from the evolution of
the density. Fig.~\ref{fig:twores} shows the evolution
of the density measured at $x=0$ for free, attractive and repulsive bosons. 
While the repulsive, and non-interacting cases are nearly identical the
attractive case shows oscillations with period $T\sim \frac{1}{c^2}$,
due to the competition between diffusion of the bosons and the action
of the attractive potential. This can be seen in the time evolution of
the wave-function for two attractive bosons,
(Fig.~\ref{fig:wfatt2}). As time evolves, the initial peaks coalesce
and increase in height. At longer times, this central peak
``breathes'' as it diffuses. The oscillations arise due to the formation of 
a bound state that appears as a contribution from a pole of the S-matrix~\cite{muth}. 
These effects are measurable in
time-of-flight experiments in both cases.  We expect the results to be
qualitatively similar for any number of particles.

Exact integrations beyond two particles are difficult. However, the
contour approach allows us to extract the asymptotic behavior of the
evolved wavefunctions. Consider repulsive bosons and carry out the
scaling $\lambda\to\lambda\sqrt{t}$. Then,
$Z^y_{ij}(\lambda_i-\lambda_j) \to \sgn(y_i-y_j)[1-
\frac{1}{c\sqrt{t}} (\lambda_i-\lambda_j) +\ldots]$, yielding to leading order,
\begin{equation*}
  \begin{split}
    \label{eq:claimt}
    &|\Phi_0,t\rangle \to
    \int_x \int_{y}\int_{\lambda}\,\theta(\vec{x})\Phi_0(\vec{x})\\
    &\times \prod_j e^{-i\lambda_j^2+i\lambda_j(y_j-x_j)/\sqrt{t}}
    \prod_{i<j} \sgn(y_i-y_j) b^\dag(y_j)\ket{0}\\
    &=
    e^{-iH^f_0 t}\int_{y}\mathcal{A}_y\theta(\vec{y})\Phi_0(\vec{y}) \prod_j
    c^\dag(y_j)\ket{0},
  \end{split}
\end{equation*}
 $c^{\dagger}(y)$ being fermionic creation operators
replacing the ``fermionized'' bosonic operators, $\prod_j c^{\dagger}(y_j)=\prod_{i<j} \sgn(y_i-y_j) b^\dag(y_j)$.
 $H^f_0=\int_x \partial c^{\dagger}(x)\partial c(x)$  is the free fermionic Hamiltonian. $\mathcal{A}_y$ is 
an anti-symmetrizer acting on $y$.
Thus, the repulsive Bose gas for \emph{any} value of $c>0$, is governed
in the long time by the $c=\infty$ hard core boson limit (or its fermionic equivalent) ~\cite{jukic,girardeau}, and the system equilibrates
with an asymptotic momentum distribution, $n_k=\langle
\tilde{\Phi}_0|c^{\dagger}_kc_k|\tilde{\Phi}_0\rangle$, determined by
the antisymmetric wavefunction $\tilde{\Phi}_0(\vec{y})=\mathcal{A}_y
\theta(\vec{y})\Phi_0(\vec{y})$ and the total energy,
$E_{\Phi_0}=\langle\Phi_0\rvert H \lvert\Phi_0\rangle$.

The  corrections to the asymptotics, can be obtained using the
stationary phase approximation to carry out the $\lambda$
integrations. We have for the repulsive case,
\begin{equation*}
    \label{eq:longtime}
    \ket{\Phi_{latt},t } \to \int_y \prod_{i<j} Z_{ij}(\xi_i-\xi_j)
    \prod_j \frac{e^{it\xi_j^2-i\xi_j x_j}}{\sqrt{4\pi it}}
    b^\dag(y_j)\ket{0}
\end{equation*}
where $\xi_y=\frac{y}{2t}$ and we drop $\frac{x}{2t}$ since the
initial conditions have finite spatial extent.
Asymptotically, the evolution of the density is not instructive, so we
compute the  $N$-boson
density-density (noise) correlation function, $\rho_2(z,z') =
\langle\rho(z)\rho(z')\rangle$. At long times,
\begin{equation*}
  \begin{split}
 &\rho_2(z,z';t) \to   \rho_2(\xi_z,\xi_{z'}) = \\
 &\frac{N^2\sigma^2}{4\pi
      t^2}e^{-(\xi_z^2+\xi_{z'}^2)\sigma^2}\Bigg[1+
    \tfrac{2}{N^2}\mathrm{Re}S(\xi_{z}-\xi_{z'})  e^{ia(\xi_z-\xi_{z'})}\\
    &\times\frac{N(1-e^{ia(\xi_z-\xi_{z'})}g_{zz'})+e^{iaN(\xi_z-\xi_{z'})}g_{zz'}^N-1}{(1-g_{zz'}e^{ia(\xi_z-\xi_{z'})})^2}\Bigg]
  \end{split}
\end{equation*}
with $g_{zz'} = 1-2c\sqrt{\pi}\sigma S(\xi_z-\xi_{z'}-ic)
  \big[e^{(c+i\xi_z)^2\sigma^2}\erfc((c+i\xi_z)\sigma)+e^{(c-i\xi_{z'})^2\sigma^2}\erfc((c-i\xi_{z'})\sigma)\big]$
and $\xi_z\equiv z/2t$.
This quantity was also studied by Lamacraft~\cite{lamacraft}
for the repulsive model. However, our result differs from his.

\begin{figure}[hbt]
  \centering \fbox{\includegraphics[width=6cm]{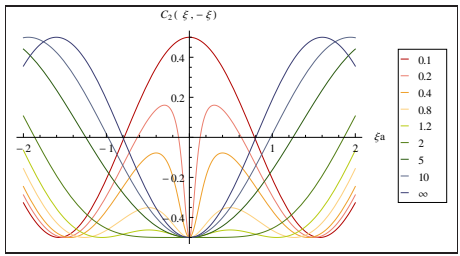}}
  \caption{Normalized noise correlation function $C_2(\xi,-\xi)$. 
    Fermionic correlations develop on a time scale $\tau\sim c^{-2}$,
    so that for any
    $c$ we get a sharp fermionic peak near $\xi=0$, i.e., at large time. 
    The key shows values of $ca$.}
  \label{fig:largec}
\end{figure}

\begin{figure}[htb]
  \centering \fbox{\includegraphics[width=2.5in]{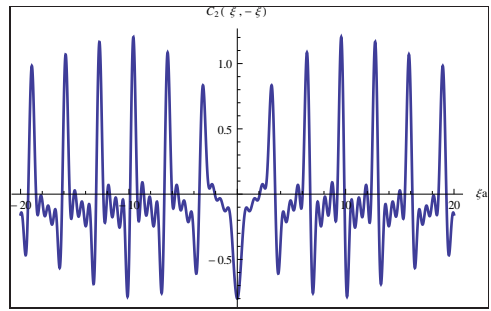}}
  \caption{Normalized noise correlation function $C_2(\xi,-\xi)$ for five repulsive bosons released from
    a Mott-like state for $ca=2$.}
  \label{fig:5prep}
\end{figure}

An interesting way to observe the evolution of repulsive bosons
to free fermions is the Hanbury-Brown Twiss (HBT)
effect.
Consider the normalized spatial noise correlations,
$\frac{\rho_2(z,z')}{\rho(z)\rho(z')} - 1\equiv C_2(z,z')$.
In the non-interacting case $S(\xi)=1$ and $g_{zz'}=1$
and we recover the HBT result for $N=2$~\cite{hbt},
$C_2^0(\xi_z,\xi_{z'}) = \frac12 \cos(a(\xi_z-\xi_{z'}))$.  At any
finite $c$, we can see a sharp fermionic dip appear (anti-correlation) that gets broader
with increasing $c$ as shown in figure~\ref{fig:largec}.
For higher particle number, as shown in fig.~\ref{fig:5prep},
we see ``interference fringes'' that get narrower and more
numerous with an increase in number of particles. However, the asymptotic fermionic
character is still visible.

\begin{figure}[bt]
  \centering \fbox{\includegraphics[width=6cm]{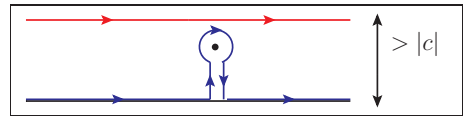}}
  \caption{Contribution from stationary phase and pole at large time
    in the attractive model. The blue curve represents the shifted
    contour.}
  \label{fig:attasymp}
\end{figure}

\begin{figure}[bt]
  \centering \fbox{\includegraphics[width=2.5in]{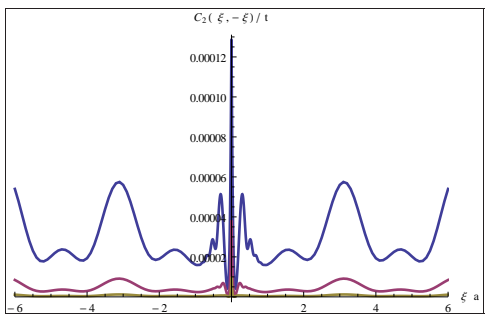}}
  \caption{$C_2(\xi,-\xi)$ for three particles in the attractive case plotted for three different times. At larger times,
the correlations away from zero fall off. $ta^2=20,40,60$ for blue, magenta and yellow respectively.}
  \label{fig:3patt}
\end{figure}

The scaling argument fails for the attractive gas as the
$\lambda$s are integrated on parallel trajectories, and as $t$ increases,
 their separation also grows. 
The asymptotic behavior of attractive
bosons, as we shall see (and already observed earlier), is dominated by a maximally bound state.
Since the contours of integration spread
out in the imaginary direction, in addition to the stationary
phase contributions at large time, we also need to take into account
the contributions from the poles.  Figure~\ref{fig:attasymp} shows an
example of how this works.
Taking into account all the pole contributions, we get a sum over
several terms given by,
\begin{equation}
  \label{eq:attasymp}
  \begin{split}
    & \ket{\Phi_{\text{latt}},t } = \int_y
    \sum_{\xi_j^*=\{\xi_j,\xi_{i<j}^*+ic\}}
    \prod_{i<j} Z_{ij}(\xi_i^*-\xi_j^*)\\
    &\times\textstyle{\prod_j}(4\pi i t)^{-1/2}
    e^{-it(\xi_j^*)^2+i\xi_j^*(2t\xi_j-x_j)}
    b^\dag(y_j)\ket{0}
  \end{split}
\end{equation}
with $Z_{ij}(-ic)\equiv c\sqrt{t}\theta(y_j>y_i)$.  Note that while
the asymptotic dynamics of the repulsive model is given purely in terms of
the ``lightcone'' variables $\xi_j\equiv\frac{y_j}{2t}$,
this is not the case in the attractive model where observables
acquire explicit time dependence. Also,
important to note is that the bound state contributions which appear
when one or more of the S-matrix poles are picked up, are higher order
in $t$ as compared to the non-bound states. Asymptotically therefore,
the larger bound states dominate. In fig.~\ref{fig:3patt} (obtained by numerically integrating
the expression for the noise correlation)
we see interference fringes similar to the repulsive case with the
central peak increasing positively  indicating
the dominance of the maximally bound state.

We now consider the evolution of the Bose gas after a quench from an
initial state where all the bosons are in a ground state of a harmonic
trap, $|\Phi_{\text{cond}}\rangle = \int_x \prod_j
\frac{1}{(\pi\sigma^2)^{\frac14}}e^{-x_j^2/\sigma^2}
b^\dag(x_j)\ket{0}$. In order to use the Yudson representation, we have to rewrite
the initial state as a sum over all different orderings of the coordinates using the symmetry 
of the initial wavefunction $\Phi_0$.
For the repulsive model, the stationary phase
contribution is all that appears:
\begin{equation}
  \begin{split}
    |\Phi_{\text{cond}}, t\rangle &= \int_{x,y} \prod_{i<j} Z^y_{ij}\left(\frac{y_i-y_j-x_i+x_j}{2t}\right)\\
    &\prod_j \frac{1}{\sqrt{2\pi
        it}}e^{i\frac{(y_j-x_j)^2}{4t}-\frac{x_j^2}{\sigma^2}}
    b^\dag(y_j)\ket{0}.
  \end{split}
\end{equation}
For the attractive case, as shown in eq.~\eqref{eq:attasymp}, the
bound states dominate, with the larger bound states (more pole
contributions) dominating at larger time. This holds independently of
the initial conditions.

\begin{figure}[htb]
  \centering \fbox{\includegraphics[width=2.5in]{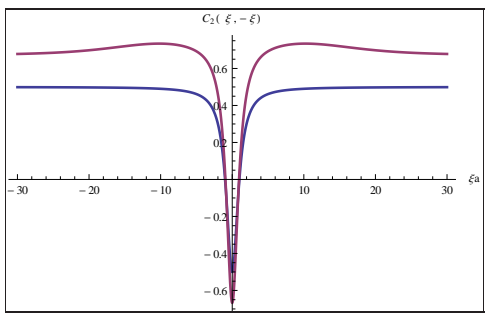}}
  \caption{$C_2(\xi,-\xi)$ for two (blue) and three repulsive bosons
    starting from a condensate. Unlike the attractive case, there is no explicit time dependence asymptotically. $ca=3$}
  \label{fig:c2repcond23}
\end{figure}

\begin{figure}[htb]
  \centering
  \fbox{\includegraphics[width=2.5in]{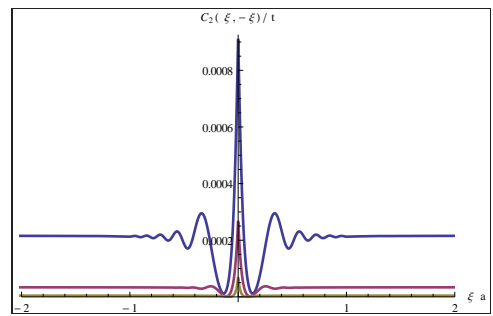}}
  \caption{$C_2(\xi,-\xi)$ for three attractive bosons starting from
    a condensate. Note that the side peak structure found in
    fig.~\ref{fig:3patt} is missing due to the initial condition. We show the evolution at three times. As time increases, 
  the oscillations near the central peak die out. Times from top to bottom $t c^2=20,40,60$}
  \label{fig:c2attcond3p}
\end{figure}

Figure~\ref{fig:c2repcond23} shows the noise correlation
for two and three repulsive bosons starting from a condensate. We
see the characteristic fermionic dip develop. The plots for three
attractive bosons are shown in fig.~\ref{fig:c2attcond3p}. The
oscillations arising from the interference of particles separated
spatially (HBT) do not appear. The attractive case however does show the
oscillations near the central peak that are visible in the case
when we start from a Mott insulator.

We conclude that although the details of the time evolution depend on
the initial state we quench from, the asymptotics show universal
features given in terms of a ``dynamic RG picture'' with the asymptotic
dynamics being controlled by $c= \pm \infty$  Hamiltonians for the
repulsive and attractive interactions, respectively, with time playing the role
of the successively reduced cut-off in an RG procedure, projecting out the effective low energy physics.
This picture suggests ``basins of attraction'' of perturbed Hamiltonians around the Lieb-Liniger Hamiltonian
(e.g., a short range potential replacing the delta function),
whose long time evolutions are given by the $c= \pm \infty$ limit.
We therefore expect our results, beyond their theoretical significance, to also provide experimental predictions.

In summary, we have used the contour integral representation
for both the attractive and repulsive Lieb-Liniger model and shown
that we can extract the asymptotic wave function and observables
analytically.  The representation overcomes some of the major
difficulties involved in using the Bethe-Ansatz to study the dynamics
of some integrable systems. The results indicate the emergence of universal dynamic features.
The application to other models is under consideration.

{\bf Acknowledgments} We are grateful to I. Klich, M. Rigol,
A. Rosch and V. Yudson for useful discussions.
This work was supported by NSF grant DMR 1006684.

\bibliography{bosegas}
\end{document}